\documentclass[onecolumn]{emulateapj} 
\usepackage{epsfig}
\usepackage{wasysym}
\usepackage[figuresleft]{rotating}
\usepackage{natbib}
\usepackage{etoolbox}
\newtoggle{emulateapj}
\toggletrue{emulateapj}
\iftoggle{emulateapj}{\newcommand{\icarus}{Icarus}}{}
\iftoggle{emulateapj}{}{}
 \iftoggle{emulateapj}{\newcommand{\Sigurdsson}{Sigur{\scriptsize \DH}sson}}{\newcommand{\Sigurdsson}{Sigur\dh sson}}

\newcommand{\msun}{\hbox{$\hbox{M}_{\odot}$}}
\newcommand{\rsun}{\hbox{$\hbox{R}_{\odot}$}}
\newcommand{\farcmin}{\ensuremath{.\!\!^{\prime}}}
\newcommand{\arcs}{\ensuremath{^{\prime\prime}}}

\newcommand{\HI}{\hbox{{\rm H}\kern 0.1em{\sc i}}}
\newcommand{\HII}{\hbox{{\rm H}\kern 0.1em{\sc ii}}}
\newcommand{\CIV}{\hbox{{\rm C}\kern 0.1em{\sc iv}}}

\newcommand{\ghat}{\^G}

\newcommand{\WISE}{{\it WISE}}

\newcommand{\kone}{\hbox{{\rm Type }\kern 0.1em{\sc i}}}
\newcommand{\ktwo}{\hbox{{\rm Type }\kern 0.1em{\sc ii}}}
\newcommand{\kthree}{\hbox{{\rm Type }\kern 0.1em{\sc iii}}}

\shorttitle{The \ghat\ search for Kardashev Civilizations I}
\shortauthors{J.\ T.\ Wright et al.}
\slugcomment{Accepted for publication in {\em The Astrophysical Journal} on 23 June 2014.}

\begin{document}


\title{The \ghat\ Infrared Search for Extraterrestrial Civilizations with Large Energy Supplies. I. Background and Justification}

\author{J.\ T.\ Wright\altaffilmark{1,2}, B.\ Mullan\altaffilmark{1,3,4}, S.\ \Sigurdsson\altaffilmark{1,2}, and M.\ S.\ Povich\altaffilmark{5}}

\altaffiltext{1}{Department of Astronomy \& Astrophysics, 525 Davey Lab, The Pennsylvania State University, University Park, PA, 16802}
\altaffiltext{2}{Center for Exoplanets and Habitable Worlds, 525 Davey Lab,  The Pennsylvania State University, University Park, PA 16802}
\altaffiltext{3}{Blue Marble Space Institution of Science, PO Box 85561, Seattle, WA 98145-1561}
\altaffiltext{4}{Carnegie Science Center, 1 Allegheny Ave., Pittsburgh 15212}
\altaffiltext{5}{Department of Physics and Astronomy, 3801 West Temple Ave, California State Polytechnic University, Pomona, CA 91768}


\begin{abstract}

We motivate the \ghat\ infrared search for extraterrestrial civilizations with large energy supplies.  We discuss some philosophical difficulties of SETI, and how communication SETI circumvents them.   We review ``Dysonian SETI'', the search for artifacts of alien civilizations, and find that it is highly complementary to traditional communication SETI; the two together might succeed where either one, alone, has not.  We discuss the argument of Hart (1975) that spacefaring life in the Milky Way should be either galaxy-spanning or non-existent, and examine a portion of his argument that we dub the ``monocultural fallacy''.  We discuss some rebuttals to Hart that invoke sustainability and predict long Galaxy colonization timescales.  We find that the {\it maximum} Galaxy colonization timescale is actually much shorter than previous work has found ($< 10^9$ yr), and that many ``sustainability'' counter-arguments to Hart's thesis suffer from the monocultural fallacy.  We extend Hart's argument to alien energy supplies, and argue that detectably large energy supplies can plausibly be expected to exist because life has potential for exponential growth until checked by resource or other limitations, and intelligence implies the ability to overcome such limitations.  As such, if Hart's thesis is correct then searches for large alien civilizations in {\it other} galaxies may be fruitful; if it is incorrect, then searches for civilizations within the Milky Way are more likely to succeed than Hart argued.  We review some past Dysonian SETI efforts, and discuss the promise of new mid-infrared surveys, such as that of \WISE.

\end{abstract}




\section{Introduction}

\label{sec:intro}

\subsection{The Question}

The fact that we have not been contacted by any extraterrestrial intelligence (ETI) is most famously encapsulated by Enrico Fermi's question ``Where is everyone?"(see \citealp{jones85} for a historical account).  Similar observations are ``Fact A'' of \citet{hart75} (``There are no intelligent beings from outer space on Earth now.'') and the apparent lack of any evidence of their communication (the ``Great Silence'' of \citet{Silence}).  

If one assumes that interstellar communication and travel are not so difficult as to be practically impossible, and one holds the Copernican\footnote{Copernican in the figurative sense that Earth is not special or unique beyond it happening to be our home.}  position that intelligent life is not unique to Earth, then one is led to an apparent contradiction.  Given the number of potential sites for intelligent life to emerge in the galaxy (of order $10^{11}$ stars) and the amount of time it has had to arise (of order a Hubble time, $10^{10}$ years), then one is led to the so-called Fermi-Hart Paradox or Problem that ETIs should be prevalent throughout the Milky Way, and yet we see no evidence of them.  

Resolutions of this problem involve contradicting its various assumptions.  The most pessimistic resolution to the paradox is that of \citet{hart75}:  we are alone in the Galaxy, or at least ETIs are sufficiently rare that none has has been inclined to engage in widespread interstellar travel.  As searches for ETIs become increasingly more sensitive, this resolution becomes increasingly favored.  More optimistically, it could be that ETIs are common, but never sufficiently spacefaring to be widespread throughout the Galaxy; or it could be that spacefaring ETIs are common, but we have not yet noticed them for some reason.

In this work we will review and reinforce Hart's basic thesis, which powerfully argues that once spacefaring ETIs arise, they should populate their galaxy on a timescale short compared to the Hubble time.  

The waste heat approach is complementary to other SETI efforts, especially communication SETI, because it makes different assumptions and can help focus them to likely targets, but might not, by itself rule out purely naturalistic phenomena.   Among the advantages to a waste heat approach is its small number of assumptions about ETI behavior and technology.  Assuming only conservation of energy, the laws of thermodynamics, and that much-faster-than-light travel is impossible, we can test for the existence of ETIs with very large energy supplies in other galaxies, out to great distances.  

This work provides background and motivates our search for alien waste heat, which we dub \ghat, the {\it Glimpsing Heat from Alien Technologies} survey (or G-HAT), and which we describe in Paper II of this series.

\pagebreak
\subsection{Plan}

We first discuss why civilizations with large energy supplies are plausible and worth searching for.  Because SETI is not a common topic in the astrophysical literature or curriculum, this paper is primarily a discussion of previous work most germane to \ghat.

In Section~\ref{philosophy} we discuss the two primary forms of SETI, communication SETI and ``artifact'' SETI, and some philosophical difficulties in searching for something whose form is uncertain.  

To illustrate that galaxy-spanning civilizations are plausible, in Section~\ref{Hart} we review Hart's argument that spacefaring life should rapidly spread throughout the Galaxy.  

In section \ref{energy} we extend this argument to the growth of an ETI's total energy supply, and show that it implies that very old, large ETI's in other galaxies should be detectable with today's mid-infrared (MIR) surveys, such as that just conducted by the {\it WISE} satellite, following the waste heat approach of \citet{dyson60} and \citet{slysh85}.   We also provide a description of the classes of intelligent spacefaring civilizations a waste heat approach would detect (which we dub a ``physicist's definition'' of intelligent life).   Ultimately, this approach will allow humanity to extend our search for ETI's (and, perhaps, Hart's ``Fact A'') beyond the lack of evidence of ETIs in the Solar System to well beyond the Local Universe.  

In Section~\ref{Maximum} we show that previous work on the {\it maximum} time it takes such a species to populate its galaxy is in error.  In fact, this is actually quite short compared to the age of the Milky Way, which reiterates and amplifies this part of Hart's argument.  

In Section~\ref{Reinforce} we apply our reasoning to some classes of rebuttals to Hart's thesis, in particular those that invoke ``sustainability'' to explain how the Milky Way could be inhabited with many spacefaring civilizations but not be saturated with them.

In Section~\ref{Dyson} we discuss the physical plausibility and detectability of Dyson spheres, and in Section~\ref{past} we discuss past searches for them.

In Section~\ref{results} we discuss the consequences of a positive and a null detection of large alien energy supplies, and in Section~\ref{conclusions} we give our conclusions to this, first, paper in the \ghat\ series.

\section{Philosophies of SETI}

\label{philosophy}

\subsection{Communication SETI}

Much work in the Search for Extraterrestrial Intelligence (SETI, see \citet{tarter01} for a review) focuses not on finding evidence of ETI travel to the Solar System, but on detecting ETI communication (either deliberate messaging or ``leaking'' of deliberate signals not intended for us).  This approach is in principle sensitive to any ETI that has achieved a level of technology similar to ours, and does not require that ETIs participate in interstellar travel; however it does require that ETIs produce signals detectable by today's technology.  It thus essentially ignores Hart's argument that ETI's would have populated the galaxy by now (neither denying nor accepting it) and seeks unambiguously engineered signals that could only be the product of an ETI.  

Directed programs in this Search for Extraterrestrial Intelligence have helped to carve out particular niches in the parameter space of ETI behaviors we can imagine. The most popular strategies or proposals for SETI have historically centered on broadcasts by an ETI within particular bands of the electromagnetic spectrum we would consider obvious or desirable (see \citealp{sagan75} for examples of recommendations). However, no ostensible signatures of ETI transmissions have been detected in wide radio bands \citep{latham93}, nor in narrow bandwidths centered on the hyperfine transition of \HI\ (\citealp{horowitz93}; \citealp{lazio02}), nor convenient multiples of this frequency (\citealp{blair92}; \citealp{lazio02}). The same is true for different atomic and molecular transitions of other substances (e.g., H$_2$O and OH: \citealp{cohen80};  CO and CN: \citealp{davidge90}; positronium: \citealp{mauer96}). \citet{siemion13} searched for narrow-band radio emission from {\it Kepler}-surveyed stellar systems with known exoplanets and extrapolated that the number of radio-loud civilizations in the Galaxy is $\lesssim$ 10$^{-6}$ \msun$^{-1}$.      
    
Alternatively, \citet{keenan99} contended that optical SETI strategies --- e.g., hunting for nanosecond pulses from MJ-scale optical lasers --- are superior. These campaigns have been thus far unsuccessful, however \citep[e.g.][]{howard04}. More exotic strategies have been proposed (e.g., \citealp{korpela08}); these include hunting for signs of artificial modification of planetary albedo, searching for intentionally generated gravity waves, and researching the utility of quantum entanglement as a communication pipeline. Others employ analyzing the CMB as a potential signal carrier \citep{kardashev79}, or hunting for collimated transmissions at mm wavelengths (\citealp{kardashev85} and references therein). These remain achievable in principle, but as of yet unrealized in practice.

However, the total parameter space searched for alien signals remains very small \citep[see, for instance,][]{papa85}, so the failure of communication SETI to date rules out only current communications of certain sorts, of certain powers, at certain duty cycles, in certain places.  Thus, today it provides very weak evidence for the non-existence in the Milky Way of electromagnetically telecommunicating civilizations (to use the term of \citet{blair92}.)

\subsection{Artifact SETI and ``Interstellar Archaeology''}      

\label{artifact}

An alternative strategy is to identify artifacts or  ``archaeological'' signatures (e.g., \citealp{carrigan12}) of ETIs \citep[somtimes called a ``Dysonian'' approach, e.g.][]{Bradbury11,vidal11}.

The most famous example of circumstellar mega-engineering is that of a ``Dyson sphere'' \citep{dyson60}, a structure or suite of structures that block a star's light (as an energy source or a living surface) and reradiate it in the thermal infrared.  We find these to be of particular interest: a civilization of sufficient technological sophistication has an obvious source of abundant free energy over $\sim$ Gyr timescales in its host star. A typical sunlike star provides $\approx$ 4 $\times$ 10$^{26}$ W throughout the course of its $\sim$ 10 Gyr lifetime (c.f.\ our current energy generation of order 10$^{13}$ W). By building an array of light-gathering structures around the star, the ETI could harness this free stellar power for its own purposes. 

Such Dyson spheres need not be single, rigid, structures\footnote{If such structures were to exist despite their apparent mechanical and gravitational unfeasibility, then presumably they would provide a surface upon which an alien civilization could exist, akin to the surface of a planet.  It is interesting to note that Earth-like conditions could be simulated around a very-low-mass star ($\sim$ 0.1 \msun) with a static sphere $\sim 1.5$ \rsun\ in radius.   Such a structure would have a surface gravity of $\sim1g$, and an effective temperature of $\sim 250$K.  It would also have a surface area $\sim 200$ times larger than the Earth and collect $\sim 100$ times the free energy from starlight; this starlight would also be available for hundreds or thousands of times as long as sunlight, thanks to the longer hydrogen-burning lifetime of very low mass stars.  At this distance, the star's magnetic field and even photon pressure might provide mechanical assistance in stabilizing the sphere's shape and position around the star.}; the actual covering fraction of a swarm of steerable energy collectors constituting the practical ``sphere" might vary, and therefore the resulting SED we receive may contain time-variable contributions from the unobscured starlight.  ``Complete'' Dyson spheres would certainly be more easily recognizable, however. \citet{sagan66} and \citet{kardashev79} showed that the IR emission of opaque Dyson spheres (with peak wavelengths 15 $\mu$m--2 mm) was detectable to $\sim$ several 100 pc by the technology of the time. 

Another approach would be to observe the effects of alien industry on its environment.  \citet{freitas85} proposed that particularly rapacious civilizations could strip solar systems of hydrogen fuel for fusion, creating circumstellar effusion clouds that are detectable in the \HI\ and tritium hyperfine transitions. Other ambitious engineering projects are theoretically detectable \citep{carrigan12}, including artificially increasing stellar lifetimes by manipulating stellar rotation, internal pressures, and/or depositing hydrogen fuel (possibly producing blue straggler stars), terraforming planets or otherwise altering a planet's atmospheric composition with known byproducts of industrial and otherwise intelligent activity.  \citet{forgan11} explored how ETI asteroid tampering could be observable in the SEDs of debris discs, and showed how such engineering projects in progress could be identifiable by anomalous chemical abundances of a mined debris disc, the thermal signature produced by waste dust, or the debris distribution itself as a dynamical signature of ETI intervention. 

Another approach is to observe products of alien mega-engineering themselves.  For instance,  \citet{Arnold05} postulated that ETIs might construct long-lived beacons in the form of planet-sized shields or louvred structures in short-period orbits around stars.\footnote{It is interesting that one of \citeauthor{Arnold05}'s speculations was apparently validated shortly after it was published.  \citeauthor{Arnold05} specifically described the {\it Kepler} mission as a prime vehicle for searches for such beacons, and gave an example of an apparent transiting planet whose depths varied by a factor of 5, with a short repetition cycle and a timing pattern indicative of a non-natural origin.  Indeed, {\it Kepler} discovered such a phenomenon in 2012 with the target star KIC 12557548 \citet{Rappaport12}.  This star shows transit depth variations of at least a factor of 6.  The discoverers interpreted this as evidence of an ``evaporating'' Mercury-sized planet, and indeed the apparently random pattern of depth variations (and lack of timing variations) distinguish the object from the sort specifically described in \citeauthor{Arnold05}'s proposal.  \citeauthor{Arnold05}'s model might be truly validated in this case if the sequence of transit depths could be shown to have some underlying pattern of unambiguously intelligent design.}  Such objects would consume little power (and, at any rate, have an abundant supply of energy nearby) but would effectively transmit simple signals across the galaxy in the form of manifestly non-natural transit light curves (i.e.\ they would have obviously non-circular aspects).  Such structures would thus also be a form of communication SETI.

\subsection{A Philosophical Challenge of SETI}

\label{challenge}

Efforts to identify known phenomena that have no natural explanation and postulate why ETIs might be the most natural cause have a major difficulty that follows from Clarke's so-called ``Third Law'' \citep{Clarke3}, which states ``Any sufficiently advanced technology is indistinguishable from magic.''   

Science, by definition, assumes that the Universe is governed by Natural Law, and seeks to interpret all observations as consequences of these laws.  ``Magic'', in Clarke's sense, means the apparent suspension of Natural Law, and so any investigation that assumes that an observation is due to a sufficiently advanced technology risks being inherently unscientific.  In particular, invoking ETIs to explain any (or all!) unexplained phenomena leads to an ``aliens of the gaps'' approach which is philosophically problematic.  Such ETI hypotheses may be difficult or impossible to disprove, and have little or no justification beyond the facts that an observation remains unexplained by natural causes, and that ETIs may exist.  

Of course, the other extreme --- assuming that all observations must be the result of purely natural phenomena are {\it not} due to advanced technology ---  is itself patently logically invalid because it {\it assumes} that ETIs have no detectable effect on the Universe.  SETI must thread this philosophical needle if it is to resolve Fermi and Hart's problem in a scientific manner.

Artifact SETI can thus proceed by seeking phenomena that appear outside the range that one would expect natural mechanisms to produce.  Such phenomena are inherently scientifically interesting, and worthy of further study by virtue of their extreme nature.  The path from the detection of a strange object to the certain discovery of alien life is then one of exclusion of all possible naturalistic origins.  While such a path might be quite long, and potentially never-ending, it may be the best we can do.

Communication SETI, on the other hand, shortcuts this path to discovery by seeking signals of such obviously engineered and intelligent origin that no naturalistic explanation could be valid.  Together, artifact and communication SETI thus provide us with complementary tools: the most suspicious targets revealed by artifact SETI provide the likeliest targets for communication SETI programs that otherwise must cast an impossibly wide net, and communication SETI might provide the conclusive evidence that an extreme but still potentially naturalistic source is in fact the product of extraterrestrial intelligence. \citep{Bradbury11}

\pagebreak
\section{Hart's Arguments For Rapid Galactic Colonization, and Rebuttals}

\label{Hart}

\citet{hart75} essentially restated the Fermi paradox, arguing that colonization of the Galaxy by an intelligent, spacefaring species should be fast compared to the age of the Galaxy.  Hart's conclusion is that since we have not encountered ETIs, or signs of them, we must be the first intelligent spacefaring species in the Milky Way.  

Hart divided objections to his argument (which are also resolutions of the Fermi-Hart paradox) into four categories:

\begin{enumerate}
\item Physical: interstellar travel is infeasible, at any level of achievable technological development
\item Sociological: extraterrestrials lack interest, motivation, or organization to visit or contact Earth.    This category includes some of what \citet{cirkovic09} dubs ``solipsist'' resolutions, in which our observations are being deliberately manipulated or confused by ETIs so as to frustrate SETI efforts.
\item Temporal: extraterrestrials have not had time to arise and reach or contact Earth yet.
\item That extraterrestrial intelligence {\it has} visited or {\it does} visit Earth, but that we have not noticed it.
\end{enumerate}

\noindent Hart dismissed solutions in each of these four categories, and concluded that the resolution to the Fermi-Hart Paradox is that ETIs do not exist in the Milky Way.

Indeed, an anthropic approach might hold that, while advanced alien life may be common in the Universe, the fact of our existence (unmolested after 10 Gyr of the Milky Way's existence and 4.5 Gyr of the Earth's) is only possible in an ``empty'' galaxy in which we are the first intelligent, spacefaring species to evolve.  This solution is consistent with both Hart's conclusion and with the premise that ETIs are common in the Universe, and could be tested if ETI's could be detected in most {\it other} galaxies.

Here, we will restate and reinforce Hart's objections to these resolutions in light of the nearly 40 years of astrophysical progress since his work.  This restatement amounts to a plausibility argument that  galaxy-spanning ETIs should exist if any spacefaring ETIs exist.

\subsection{Physical Arguments}

While Hart had to argue that interstellar probes were well within the realm of technical feasibility, barred by neither time nor energy consumption limitations, we can do better today. {\it Voyager} 2 and both {\it Pioneer} probes have traveled beyond the orbit of Neptune, and once they and {\it New Horizons} join {\it Voyager 2} in crossing the heliopause they will all properly be called interstellar \citep{Kerr2013}.  At their current speeds and distances from the Sun, they will require of order $10^5$ years to reach distances comparable to the nearest stars (although they are not traveling towards any of them).  Although this time is long, it demonstrates that interstellar travel by human-built probes is certainly possible.  

These probes would certainly no longer have functioning power supplies by the time they might arrive at an alien planetary system, but this is a function only of their engineering.  Hart points out, correctly, that nuclear energy could easily provide not just sufficient operational energy, but indeed thrust for deceleration for more advanced probes.

The velocities of our existing interplanetary probes are large (of order $10^{-4} c$), but there is no engineering or physical reason why they could not be at least an order of magnitude larger.  Given the possibility of centuries, millennia, millions of years, or hundreds of millions of years of future human technological advancement, it is clear that these spacecraft represent an extreme lower limit to the speed and utility of interstellar probes.  

A more ambitious approach would be to send colony ships to other star systems; this additional complication adds considerable engineering difficulty but, again, no fundamental physical obstacle, and indeed advances in biology and medicine in the last 40 years have made such approaches more plausible.  Whether the ships would, for the long cruise times of their journeys, contain conscious colonists; humans in some sort of ``suspended animation''; or merely the machinery for gestating and raising humans from embryos, gametes, or even just genetic material; varies only the degree of complexity of the problem.  The fact that we can today imagine and even design ships that could do this implies that the barriers to their construction are a matter of will and achievable technological development, not fundamental limits of physics or engineering.

\subsection{``Sociological'' Resolutions}

\subsubsection{The ``Monocultural Fallacy''}

\label{monocultural}

In light of the many of the proffered resolutions to the Fermi-Hart Paradox in the past 40 years, Hart's most underappreciated argument must be the one involving the so-called sociological explanations, and so it bears repeating.  Such explanations invoke a typical behavior pattern of alien species (e.g.\ perhaps they: do not wish to travel / do not wish to  communicate / destroy their climate / destroy themselves with nuclear weapons / do not allow ``immature'' species to ``join the interstellar club'' / achieve transcendence / become immersed in computer simulations, etc.).  Hart rightly argues that these must fail because they necessarily must apply to every alien species in the galaxy, and must apply {\it for the entire history of their existence}.  Or, put more elegantly by Hart, any such solution 
\begin{quote}
...might be a perfectly adequate explanation of why, in the year 600,000 {\sc bc}, the inhabitants of Vega III chose not to visit the Earth.  However, as we know, civilizations and cultures change.  The Vegans of 599,000 {\sc bc} could well be... more interested in space travel.  A similar possibility would exist in 598,000 {\sc bc}, and so forth.  Even if we assume that the Vegans' social and political structure is so rigid that no changes occur even over hundreds of thousands of years, or that their basic psychological makeup is such that they always remain uninterested in space travel, there is still a problem... [Such a solution] still would not explain why the civilizations which developed on Procyon IV, Sirius II, and Altair IV have also failed to come here.  [Such a solution] is not sufficient to explain [the apparent absence of aliens] unless we assume that it will hold for {\it every} race of extraterrestrials --- regardless of its biological, psychological, social, or political structure --- and at {\it every} stage in their history after they achieve the ability to engage in space travel.  That assumption is not plausible, however, so [such solutions] must be rejected as insufficient. \citet{hart75}
\end{quote}

We dub this the ``monocultural fallacy'', that considers a solution that could plausibly apply to a single culture of a single alien species, and attempts to apply it to every culture of every alien species across the breadth {\it and  history} of the galaxy, without exception.  Saving a sociological explanation by asserting that spacefaring alien species have only arisen a few times (and so the explanation need be only applied a small number of times, making it more plausible) is tantamount to asserting that intelligent alien life is rare, consistent with Hart's conclusions.

\subsubsection{Sustainability} 

\label{sustainability}

One version of a ``sociological'' solution that may seem to evade part of Hart's critique involves inevitable self-destruction; that is, that the technologies and coordination that allow spacefaring also allow fantastically destructive technologies and the possibility of catastrophic cultural collapse, either or both of which inevitably lead to a species' extinction \citep[see, e.g.][for one sober account]{vonHoerner75}.  This is most famously encapsulated in the final term ``L'' in many versions of the Drake Equation \citep{DrakeEq,DrakeL}, although that term is most appropriately interpreted as specifying the average {\it communication} timespan of a civilization.  It thus includes the possibility not only of the complete destruction of a civilization but some sort of technological transcendence that renders them effectively mute or invisible to our detection

Examples of the argument that exponential growth inevitably triggers the collapse of civilization include the ``light cage limit'' of \citet{McInnes02} and  the ``Sustainability Solution'' of \citet{haqq09}).  We refer to the class of solutions that begin with the assertion that ``exponential growth is unsustainable'' before the galaxy is colonized as adopting the ``sustainability hypothesis.''

The Light Cage Limit comes from the mathematical impossibility of exponentially increasing resource use with short timescales.  \citet{McInnes02} imagines that sustained resource growth would require an expanding sphere within which to gather resources, and this sphere must eventually expand faster than light to sustain exponential growth.  Hitting this ``Light Cage Limit'' then triggers one of two extreme results: catastrophic social collapse, causing the civilization to go extinct, or else a coordinated, voluntary check on growth that slows population growth to permanently sustainable levels.  In either case, the consequence (it is presumed) is cessation of galactic colonization, thus explaining Fact A.

The Sustainability Solution similarly, but less specifically, asserts that since exponential growth is unsustainable, there can only be two sorts of civilizations:  slow- or non-growing, (potentially) long-lived ones, and fast-growing, necessarily short-lived ones.  Both are incapable of sustained exponential colonization because they are, by definition, non-exponential and unsustained, respectively.

We discuss our rebuttal to the sustainability hypothesis in Section~\ref{rebuttal}.

\subsection{Temporal Arguments}

Given the vastness of the distance between the stars, and the apparent technological implausibility of sending a probe to even the nearest star within the span of even a long-lived human, it may seem absurd to consider a galaxy-spanning supercivilization plausible, but the timescales for galactic colonization are actually quite short compared to the age of the Galaxy, even for civilizations with ``slow'' travel capabilities.   Indeed, optimistic assumptions lead to very short {\it minimum} colonization times: for instance, \citet{cirkovic09} and \citet{tipler80} calculated that galactic colonization should only take $\mathcal{O}$(10$^6$) years, and \citet{Armstrong13} argue that entire galaxy groups can be populated in much less than a Hubble time.

There are two primary timescales to consider: that for the first civilizations to arise in a galaxy, and then for colonization of the galaxy.  

\subsubsection{Timescale To the Rise of Civilization}

On the first of these, it is possible that spacefaring civilizations have only recently arisen, and so have not had time to colonize the Galaxy, implying that we are among the first spacefaring civilizations in the Milky Way.  Evidence now indicates that terrestrial planets could have formed very close to the formation of the Galaxy itself \citep{buchhave12}. With some insight into stellar chemical evolution, the timescales needed to develop biological complexity, and the time-dependent rate of lethal supernovae events, \citet{GHZ} placed the oldest possible extrasolar habitats forming $\approx$ 8 Gyr ago, in the circumgalactic region known as the Galactic Habitable Zone (\citet{gonzalez01}, but see \citet{Prantzos08}). \citet{livio90} and \citet{livio99} inferred that there must exist a monotonic, positive relationship between the timescales of stellar lifetimes and noogenesis. Had these two timescales been independent, it would be statistically more likely that ETI is very rare. Instead, ETI could have emerged when pertinent elements for organic life were in sufficient quantity, which the authors similarly traced to $\approx$ 10 Gyr ago.   

On the other hand, \citet{cirkovic08} and \citet{annis99a} have argued that a phase-transition model of galactic habitability that would explain the ``Great Silence'' in light of the extreme age of the Galaxy.  For instance, GRBs tend to occur on timescales similar to the $\sim 10^8$ yr colonization time of the Galaxy, and might sterilize large swaths of the Galaxy too often for widespread colonization can occur.  Such effects could conspire to keep the development of life or intelligence stalled, preventing the development of any civilizations much more advanced than ours.  Because GRBs proceed from the production (and destruction) of high-mass stars, and these require high star formation rates, GRBs should have been more common when the Galaxy was young and its star formation rate was higher. This implies that the GRB rate should decline as the Galaxy's star formation rate declines (at least for ``long'' GRBs). Therefore, we might expect to be in an astrobiological phase transition as the rate of externally triggered catastrophe declines and the universe becomes less hostile to spacefaring civilizations.   Such a transition might occur at a common time for all galaxies in the universe.  

But absent such a phase transition, we should expect that if life is common, it arose many billions of years ago.

\subsubsection{Timescale for Galactic Colonization}

According to models of galactic colonization employing percolation theory (\citealp{hair13}; \citealp{landis98}), expanding colonies can be hemmed in by stationary ones, allowing kpc-scale voids of space empty of ETI activity that allow the unfettered development of more primitive civilizations like ours. Similar work with probabilistic cellular automata \citep{vukotic12} shows strong clustering of advanced civilizations and their colonies, with large portions of the Galactic Habitable Zone unoccupied \citep{GHZ}. These groups raise the interesting possibility that the engineering efforts of ETI exist, but have yet to be unambiguously identified from our position in a more auspiciously rural stellar neighborhood.


\citet{newman81}, modeling galactic colonization and the population pressure that encourages it as an interstellar diffusion problem, concluded that the commonly accepted Hart colonization timescale of $\sim 10^8$ years is accurate only for long-lived civilizations experiencing considerable and consistent population growth. If the civilization experiences zero population growth, the colonization time can stretch to $\sim$ 10$^{10}$ years (see also \citealp{forgan12} for a zero-population growth perspective on interstellar flights using stellar slingshot mechanics). 

\citet{carrigan12} suggested that localized ``Fermi bubbles" of expanding ETI fronts could be seen, perhaps as optical voids or strongly radiating features in infrared.  However, as Carrigan noted, bubble structures are fairly common phenomenology in extragalactic astronomy (e.g., in flocculent spirals), and Galactic astronomy \citep[e.g., IR bubbles][]{Churchwell06,Simpson12} so any artificial varieties could be difficult to spot. Galaxy arm widths tend to be on the order of $\sim$ kpc, so it is difficult to identify structures below this size scale.  Annis \citep[cited in][]{carrigan12} recommended searching for these Fermi bubbles in elliptical galaxies to avoid confusion with natural spiral structures.

We discuss problems with these calculations in Sections~\ref{fastcolony} \& \ref{nobubble}.

\subsection{They Exist and We Have Not Noticed Them}

Hart's original argument centered on his ``Fact A,'' that we have no evidence of ETI's here in the Solar System {\em today}.  That is, that the Solar System has not yet been colonized.  His fourth class of resolutions to the Paradox, ``Perhaps They Have Come'', is that they have colonized the Solar System (or, at least, visited it) but we have not noticed.  Hart assumed that it is unlikely that the Solar System has been visited but not colonized:  in the history of the Galaxy it would only take one passing colony ship to permanently inhabit the Solar System (or, at least, permanently scar it with its civilization), and given the timescales involved we should have been visited many times over in the history of the Solar System.   Indeed, it is hard to imagine how we might miss evidence of ETI colonization of the Solar System, even if they happened to go extinct before present day \citep[but see, for instance][]{Haqq12}.  

\section{The Plausibility of Very Large Energy Supplies}

\label{energy}

Hart's conclusions are that we are alone in the Milky Way and that, provided we survive as a species long enough to begin space colonization, we will ``probably occupy most of the habitable planets in our Galaxy''.   That is, galaxy-spanning supercivilizations are inevitable if intelligent spacefaring life is.    Much of Hart's reasoning extends beyond just {\em travel} across a galaxy into {\em energy supply\footnote{Energy {\em supply} is jargon that refers to the total energy generated or collected in a unit time (usually per annum); it thus represents a {\it power}, not the total quantity of energy available for all time.  It is distinguished from energy {\em use }, which does not include losses such as those associated with conversion to or transmission of, for instance, electricity.  For considerations of waste heat, it is appropriate to consider the supply, even if the actual energy use of a civilization is significantly smaller.} growth} as well.  In this section we argue that, not only are galaxy-spanning civilizations plausible, but so is our ultimate detection of their waste heat.

Because they reproduce, species of life are, in principle, capable of exponential growth until something checks them.  Mold spreads to the edge of its petri dish; vegetation grows and thickens until no sunlight reaches the forest floor; rabbits' numbers grow until they enter a chaotic dynamic with their predators.  Certain classes of intelligent, spacefaring life can achieve the sustained growth required to consume a significant fraction of the starlight in a galaxy because some classes of intelligence (those satisfying our ``physicist's definition'') implies the ability to overcome the Malthusian limitations that constrain non-intelligent life.

In what follows we are deliberately outlining only the likely {\it physical} limitations to the growth of intelligent life and its energy supply, and downplaying {\it biological} limitations exhibited by Earth life.  We do this not because we are unaware of the biological limitations of terrestrial life, but because we seek to establish the maximum possible size of an energy supply (or lack thereof), and because by our ``physicist's definition'' of intelligence ETI's can, in principle, transcend those limitations and approach physical ones.   

These definitions are not intended to ``define away'' ETIs that do not have large waste heat signatures as ``unintelligent'', but rather to define the sorts of ETIs that a waste heat search is sensitive to.

\subsection{A ``Physicist's Definition'' of Life and Intelligence}

Waste heat searches may be sensitive to what we dub to be a ``physicist's definition'' of life:  matter that processes resources and energy to produce more of itself.  We acknowledge that this definition is so broad as to potentially include abiological self-replicating phenomena such as crystals, but it needs to be broad so that we do not exclude exotic forms of life that might exist in the universe that we merely have not yet considered.  Despite this, even this definition may not be broad enough to encompass all forms of life in the universe.

Our ``physicist's definition'' of {\em intelligent} life is life that can overcome local resource limitations through the application of energy.  This roughly tracks with our intuitive understanding of intelligence.   A barnacle on a whale has no volition or ability to acquire food; it can only instinctively collect food that comes to it.  A whale, on the other hand, can expend energy to hunt for additional food.  Octopuses can apply torque to a lid of a jar to extract the food within; birds can drop shells to break them and acquire the food inside; humans can desalinate sea water and fertilize crops to create new sources of fresh water and food.  

Taken literally and applied over long timescales, this definition of intelligence might be so rough as to encompass non-intelligent behavior, especially if applied to a group of organisms as a whole (such as ant colonies, or even a grove of trees that spreads to cover a hillside, thus gathering additional solar flux), and it can be unclear to what degree instinctive behavior in an individual can be properly called intelligence.  

But of course, intelligence is not a binary proposition:  humanity's capacity for overcoming local resource limitations far exceeds that of cetaceans, and certainly dwarfs that of trees. So despite their lack of rigor, these definitions are useful from the perspective of the search for intelligent life in the universe.  If an intelligent species is spacefaring, then it will by definition consume resources with the possibility of exponential growth, and it can overcome a local lack of resources by acquiring additional energy by leaving its home planet (or other place of origin).   

We argue that if a species is spacefaring, then its level of intelligence is such that there is no practical resource limitation that it cannot overcome, except that of energy.  On Earth, apparent limitations of arable land, freshwater, and energy have been consistently pushed back as technology has advanced.  Desalinization and hydroponics illustrate how, in principle, our food supply is limited only by our energy supply and our will to apply it.   While it is true that the Earth's population may stabilize in the near- or medium-term future for sociological reasons, this does not necessarily have any implications for the long-term future of the species as it expands into space.

For brevity and specificity, we will use these definitions of ``intelligence'' and ``life'' for the rest of this paper, but we acknowledge that there may be other forms of intelligence in the Universe not captured by our ``physicist's definition.''  Naturally, a waste heat approach will not be sensitive to such life.

\subsection{Free-Energy Limited Species}

In the limit that a species or civilization uses its entire local energy supply, we can refer to it as a free-energy limited species.  ``Free energy'' here refers to the amount of work that can be extracted from an energy source, when that work is performed at a given operating or ambient temperature.  Since energy must be conserved, all energy used to do this work must be expelled, usually as waste heat, with a higher entropy than it had before it performed the work, unless it is stored.\footnote{Waste heat, by definition, cannot be used to do work in an environment at or warmer than the heat's temperature.  As we will see, this will set a practical lower limit to the temperature of waste heat of ETIs.}

There are many definitions of free energy, reflecting the complication that the precise amount of work that can be done with a given energy source will depend on the nature of the mechanism doing the work, and how it changes its environment.  Because we are making order-of-magnitude argument, and to avoid committing ourselves to particular models of alien technology, we will use the concept generally.

The term ``energy limited life'' is usually applied to parts Earth's biosphere with barely enough free energy to sustain any terrestrial life at all, and so what life there is has its metabolic rate limited by free energy, not some other resource.  

Intelligent life, being able to apply energy to overcome every other resource limitation, has the potential to become free-energy limited on a much larger scale.  Such a species could still grow its energy {\it use} (by increasing its efficiency), and it may be able to expand to acquire additional free energy (from other stars, for instance), but we do not expect to be able to detect much low entropy energy from it (because it is using as much as it can), and we should expect to detect nearly its entire energy supply as it is reradiated as waste heat.

It is perhaps likely that no species can ever become truly free-energy limited, but any species that approaches this limit (say, consuming more than a few percent of the available free-energy) will have a marked and detectable effect on the MIR luminosity of its host planet, star, or galaxy.  As we will see, a truly free-energy limited species would have a profound and easily-detected effect on its host star(s), and a thorough waste-heat search with currently available data can rule out certain classes of free-energy limited species out to intergalactic distances.

\subsection{Terrestrial Limitations on Biological Energy Use}

\label{physicists}

Life on Earth has spread nearly everywhere.  A majority of its surface has life, so a test for metabolic processes would succeed on most randomly selected samples of Earth's soil or water (in some cases even the ice, air, and deep Earth would pass such a test).  The sunlight that lands on Earth is used directly by vegetation across the globe (and indirectly by organisms higher on the food chain), and even more of this energy is used passively by life because it warms the environment and drives the wind, weather, and climate upon which many species rely.


Obviously, exponential growth is limited in individual species by a variety of factors, but the biosphere considered as a whole has managed to expand the amount of solar energy captured for metabolism to around 5\% (Paper II), limited by the nonuniform presence of key nutrients across the Earth's surface --- primarily fresh water, phosphorus, and nitrogen.  Life on Earth is not free-energy-limited because, up until recently, it has not had the intelligence and mega-engineering to distribute Earth's resources to all of the places solar energy happens to fall, and so it is, in most places, nutrient-limited \citep[see, for example discussions in][]{Reich06,Penuelas11}.

An intelligent species, by our ``physicist's definition'', could apply some of the unused solar energy to redistribute these limiting resources, and so expand life's footprint beyond what 4 billion years of evolution has managed.  Indeed, artificial fertilization and irrigation are exactly this redistribution, and we rightfully identify the development of these methods as a turning point in human history and a hallmark of civilization.  Today, humanity has made photosynthetic activity common in many nominally inhospitable parts of the American Southwest and Middle East, for instance.  If we consider our own energy generation as part of the biosphere's metabolism, then photovoltaic arrays, wind farms, and other forms of solar power stations represent further extensions of this trend.  

Non-intelligent life on a planetary scale is not directly detectable from its waste heat, because the ground beneath it would absorb and reradiate thermal energy even in the absence of life.   But spacefaring intelligent civilizations, by our definition of them, are able to extend beyond the spatial limits where planets naturally re-radiate starlight to their circumstellar and pan-galactic environments, and so are susceptible to searches for their waste heat.  


\subsection{Humanity as Case Study}

Humanity's ever-increasing energy demands drive the exploitation of more and more sources of energy with higher and higher degrees of difficulty.  Even if one country, today, cannot increase its energy supply, or simply declines to increase it, this does not mean that humanity as a species will {\it forever} cease to increase it.  Indeed, a major difficulty today is not how humanity might continue to increase its energy supply, but how to manage sufficient international cooperation even to {\it limit} its fossil-fuel consumption in the face of ecological catastrophe.

It is instructive to note along these lines that humanity's energy supply has doubled in the past 30 years.  Left unchecked, such a doubling rate would imply an energy supply equal to the total incident solar flux on the Earth in  $\sim 400$ years, at which point direct heating of the Earth would create a significant increase in temperature.  If this energy generation were accomplished entirely from non-Solar sources then the equilibrium temperature of the Earth would increase by $\sim 50$ K, rendering it essentially uninhabitable.  If accomplished from solar sources, the interception of such a large fraction of sunlight would have dramatic consequences for climate and the biosphere.

From this we can conclude that if no other limits or global cooperation intervene, humanity's energy supply on Earth must maximize at some point in the next few centuries, having effectively saturated the planet's capacity for waste heat.  But it would be incorrect to conclude that this somehow invalidates the conclusion that energy supply growth may be inevitable.  On the contrary, {\em this calculation illustrates how rapidly, on a cosmic timescale, humanity's energy needs can approach fundamental limits} (see, e.g. \citep{vonHoerner75}).  As humanity expands into space, these limits will expand dramatically, and as humanity's technology level increases its ability to approach them will only improve.

\section{Reinforcing Hart's Temporal Argument}

\label{Maximum}

Having established Hart's argument for colonization, and extended his reasoning to energy use, we now consider some rebuttals to Hart's argument, and add our own observations.

\subsection{Maximum Timescale for Galactic Colonization}


As we have discussed, percolation and automata theory has been a popular approach for estimating the timescale for Galactic colonization \citep{hair13,landis98,vukotic12}.  In these approaches, one typically considers a static network of stars, and so many of these models do not account for Galactic shear and the physics of halo orbits in their calculations.  In particular, \citet{newman81} found that colonization times could reach $\sim 10^{10}$ years for slowly growing civilizations because they were limited to expand along their frontiers, and that they eventually ``outgrow'' their colonization phase (the latter is also an example of the monocultural fallacy).

The static model of stars, in which a supercivilization can be said to occupy a compact and contiguous region of space, is a reasonable approximation for short times ($\lesssim 10^5$ years) and in the case of fast ships (with velocities in significant excess of the typical thermal or orbital velocities of the stars, so $\sim 10^{-2} c$).  In such cases, the stars essentially sit still while the ships move at a significant fraction of $c$ and populate a small region of the galaxy in some small multiple of the region's light-crossing time.

But for longer times or slower ships, the model fails badly because the stars in the galaxy are not static, even in a rotating Galactic frame.  

\subsection{A Model for the Inevitably Fast Colonization of the Galaxy with Slow Spacecraft}

\label{fastcolony}

Let us consider conservative timescales involved for a ``slow'' civilization in the Milky Way.  In the interest of calculating a {\it maximum} galaxy colonization time, we will assume a single spacefaring civilization arises, and that it is confined to launching colony ships traveling at $10^{-4} c$ (i.e.\ comparable to the orbital speeds of the planets in the Solar System, and to that of our interplanetary probes).  To be further conservative will assume zero technological development for the homeworld and the diaspora after the launch of the first ship. Finally, we will further conservatively assume that such ships are launched at a very slow rate, every $10^4$ years, and have a maximum range of 10 pc.

Travel time to the nearest stars is then of order $ 10^{-4} c / 1\mbox{pc} \sim 10^5$ years, in which time $\sim 10$ ships will be launched.  Since this travel speed is also comparable to the velocity dispersion of stars in the Galactic midplane, this timescale also describes the time it takes for stars to mix locally, bringing new stars into range of the colony ships.  To first order, the stellar system can thus continue to populate the 10 nearest stars every $10^5$ years, without immediately saturating its neighborhood with colonies or the need to launch faster or longer-lived colony ships to continue its expansion.  Further, arrival of the colony ships at the nearby stars should not be modeled as a pause in the expansion of the civilization.  Rather, the colonies themselves will continue to travel at these speeds with respect to the home stellar system, and themselves encounter fresh stars for colonization every $10^5$ years, during which time they can also launch 10 colony ships.  

Since the time to launch a new colony ship in our model is shorter than the cruise times, the expansion rate of the supercivilization is set by the velocity dispersion of the stars, not the speed of the ships or the time it takes a colony to generate a new ship.  Halo stars further shorten galactic colonization timescales because of their high relative velocities to the disk (of order $10^{-3} c$).  Halo stars, being significantly more metal poor than the disk, may have too few planets or other resources to make effective colonization destinations, but they can still be used to provide gravity assists to colony ships, giving them speeds relative to the local standard of rest of $10^{-3}c$  .  Thus, the travel time to any point in the galaxy for a ship is the travel time to the nearest suitable halo star ($10^5$ years in our example), plus the length of a cruise phase at $\sim 10^{-3} c$, plus the time required for deceleration (also $10^5$ years, if another halo star near the destination is used).  This scheme would allow the ships in our example to extend their range to 100 pc.  

If we drop the assumption of a maximum range for our colony ships, halo stars provide a capability to travel 1 kpc in 3 Myr, dominated by the long cruise time.   Since the typical velocity of halo stars with respect to a disk star is simply the orbital speed of stars at a given galactocentric radius, use of halo stars for gravitational assists allows a ship to cross the galaxy on a rotational timescale, by definition. 

The slow expansion of an ETI should thus be modeled not as an expanding circle or sphere, subject to saturation and ``fronts'' of slower-expanding components of the supercivilization.  A better model is as the mixing of a gas, as every colonizing world populates the stars that come near it, and those stars disperse into the galaxy in random directions, further ``contaminating'' every star they come near.  If halo stars are themselves colonizable, then those that counter-rotate and remain near the plane will provide even faster means of colonization, since they will encounter $\sim 10$ times as many stars per unit time as disk stars.    Non-circular orbits also provide significant radial mixing, and Galactic shear provides an additional source of mixing that is comparable to that of the velocity dispersion of the disk stars once the colonies have spread to $v_{\rm rot} / \sigma_v \sim 1/10$ of galaxy's size, or $\sim 1$ kpc from the home stellar system.  

We conclude that once a civilization develops and employs the technology to colonize the nearest stars, it will populate the entire galaxy in no more than $\sim 10^{8} $--$ 10^{9}$ years, given our deliberately conservative assumptions of colony ship launch rate ($10^{2}$ Myr$^{-1}$), cruising speeds (30 km s$^{-1}$), and technological advancement rate (zero).  More ``realistic'' values for these parameters will only decrease the galatic colonization timescale.  

\subsection{Implications For Extragalactic SETI}

\label{nobubble}

Thus, we see that galactic colonization timescales are likely to be at least one orders of magnitude shorter than the ages of galaxies, and rotational shear and the thermal motions will disperse and ``mix'' any Fermi bubbles on a rotational timescale.  Indeed, we find that the maximum timescale for Galactic colonization for a spacefaring civilization is on the order of a Galactic rotation ( $10^8$ years), {\it even for present-day probe speeds}.

It is therefore only for a brief period of a galaxy's history that an ETI would have populated only single contiguous part of a galaxy, and we should expect only a brief transition period between the rise of the first spacefaring civilization(s) and a galaxy-spanning supercivilization.  To first order, one would expect the fraction of galaxies that have been only partially populated to be the ratio of the colonization timescale to the galaxy's age, or roughly $10^8/10^{10} \sim 1\%$.  Until we have discovered 100 galaxy-spanning supercivilizations, we should not expect to find any Fermi bubbles \citep{Papagiannis80}.

There are important implications for extragalactic waste heat SETI, as well.  Because stars in a galaxy will thoroughly ``mix'' a civilization with their random motions, especially in an elliptical galaxy where there are fewer strongly correlated velocities, the waste heat of an alien supercivilization should be smoothly distributed across the galaxy.  This is in contrast to interstellar dust, which, being self-gravitating and dissipative, will tend to clump and collect into a disk.  Morphology could thus be a powerful discriminant between midinfrared emission from dust and ETI's \citep{CarriganAstronomy}.

\section{Reinforcing Hart's Sociological Argument}

\label{Reinforce}

Our fast {\it maximum} colonization timescale for the Galaxy allows us to reconsider some rebuttals to Hart's sociological argument.

\subsection{Rebutting Extinction Arguments}

In the case of the possibility of extinction, Hart's temporal reasoning appears at first to work against him:  while the Earthlings of the 20th century may have avoided global thermonuclear war, this does not mean that the Earthlings of the 21st century will avoid extinction by a manufactured supervirus plague, or that those of the 22nd century will avoid catastrophic climate change from runaway greenhouse warming, or that those of the 23rd century will avoid becoming fodder for runaway, exponentially self-replicating nanobots.\footnote{``Grey goo'' (see ``Some Limits to Global Ecophagy by Biovorous Nanoreplicators, with Public Policy Recommendations'' Freitas (2000): http://www.foresight.org/nano/Ecophagy.html).\label{goo}}  In other words, even for a fixed extinction probability per unit time, extinction is inevitable, and it is reasonable to suppose that extinction probability per unit time would grow (or at least not decrease) with technological development.

This argument has merit, but it only applies during the narrow time frame during which a civilization has the capability and propensity to destroy itself {\it and all of its colonies}.  This certainly applies to Earth today, but once a self-sufficient lunar or Martian colony exists, the colony can act as a ``lifeboat'' that can repopulate the Earth after the effects of any disaster.  Once nearby star systems are colonized, humanity's expansion into the Galaxy can proceed even without the Earth, even if colonies occasionally extinguish themselves.   As long as the periods between extinction are typically longer than the periods between the launch of successful colony ships, the expansion of life into the galaxy can proceed apace.

Thus, the existential threats to any spacefaring species, including both externally and internally induced extinction or cultural collapse, decrease with its size.  When confined to a single planet, many dangers exist, but a single self-sufficient colony will immunize the species from intra-planetary war, climate change, and even asteroid or comet impact.  Spreading to a nearby star eliminates extinction from interplanetary warfare or a catastrophe with its host star.  Once the civilization grows across a significant fraction of the Galaxy, either from fast colony ships or because the host stars have spread apart naturally, supernovae and gamma-ray bursts (GRBs) will also fail to be effective sterilizers.    

This defense is quite robust: even turning the same mechanisms that make the spread and persistence of spacefaring life robust against life does not defeat this line of reasoning.  For instance, a fleet of self-replicating machines \citep[``von Neumann'' machines,][]{vonNeumann2,vonNeumann} that attempted to exploit the powers of exponential growth could rapidly colonize the Galaxy to perform some task.  But even if this task were to scour the galaxy of life \citep[``berserkers'', e.g.][]{Berserker} they would eventually become ineffective, since the technology of colonies on the far side of the galaxy will have advanced by millions of years by the time the berserkers arrive.  Even if one invokes berserkers that can grow and adapt to such advancement and succeed in their task, the berserkers {\it themselves} would constitute an intelligent, spacefaring, galaxy-spanning supercivilization by our definition of it, which defeats the purpose of the argument.

\subsection{Stellar Distances Preclude Interstellar Coordination}

The monocultural fallacy includes treating a single species as a single culture.  Even if a civilization is not generally inclined to colonize or communicate, this does not mean that it will not happen.   To use a human analogy, it was certainly true that ``European culture'' before the 15th century had little or no inclination to cross the Atlantic and colonize the New World, but this did not prevent a sub-culture within Europe from accomplishing the task (twice!).  

Indeed, the number of distinct civilizations within a species will increase as the size of the civilization grows.  This is because ``culture'' is a phenomenon born of coordination and collective action, which become more difficult when communication becomes slow, or the possibility of coercion becomes remote because of the distances involved.  To continue the European analogy, long communication times and very slow exchange of population meant the colonies of the European powers had distinct cultures, which in some cases were kept in line only through the great expense of extending military force over long distances.  

While radio communication can, in this sense, make a planet more culturally integrated, and so potentially more uniform, interplanetary and interstellar distances make even this solution inadequate.  Once colonies are separated by light travel times that are a substantial fraction of a lifetime or significantly longer than the timescales over which cultures evolve, the coordination and collective action that characterizes a single culture become impossible.  An interstellar Napoleon could not hope to bring under his sway colonies beyond a radius of one light-lifetime.

We should then anticipate that extraterrestrial civilizations that span more than one planetary system should not be characterized by a single set of cultural characteristics, but as a supercivilization.  We use the term ``supercivilization'' to be the set of a large number of widely separated civilizations with distinct cultures, whose shared features result from their common origin, not any social coordination (which would be impossible due to the large light travel time involved).  Supercivilizations are thus resistent to many sociological resolutions to the Fermi-Hart Paradox.  

Since supercivilizations are the consequence of large light travel time, faster-than-light communication (or travel) would make it possible to avoid cultural diversity.  Indeed, if travel and communication time can be made arbitrarily short (a convenience adopted, for instance, in much science fiction), then a highly organized species might be able to impose a single culture across a large region of space;  one could rule the Galaxy (or the universe!) from the White House.  The spatial portion of the monocultural fallacy, then, is subject to the (reasonable) assumption that the speed of light is an inviolable maximum.  The temporal aspect, similarly, is subject to the assumption that (backwards) time travel is not possible. 

Thus, we should not expect spacefaring civilizations to go extinct (even from external causes!) once they form supercivilizations so large that the timescale for communication across the supercivilization is larger than the timescale for technological advancement and cultural evolution.  

\pagebreak
\subsection{Rebutting Sustainability Arguments}

\label{rebuttal}

\subsubsection{Lack of Universal Mechanism}

In the case of the sustainability hypothesis (Section~\ref{sustainability}), it is certainly reasonable to suggest that individual long-lived civilizations must inevitably stabilize their population (and, implicitly, their energy use).  This certainly seems to be true of humanity in the short-run future, as standards of living improve globally and birth rates decline.  Alternatively, such a stabilization might come about because of a limitation of resources, such as land or some element key to that civilizations' technology, or negative feedback from catastrophic climate change induced by greenhouse warming or direct heating.  

Nonetheless, any proffered solution to the Fermi Paradox based upon the sustainability hypothesis must address three issues.  First, it should specify at least one plausible mechanism by which rapidly growing civilizations inevitably collapse (since, for instance, the collapse of innumerable civilizations on earth has not yet permanently checked our species' exponential population and energy growth).  Secondly, it should explain why these mechanisms prevent not just exponential population growth within the civilization, but galactic colonization by that civilization (e.g.\ why it precludes a galactic super-civilization composed of sustainable civilizations).  Finally, it should explain why such a mechanism would be universal not just across species, but within species across both space and time (i.e.\ why it is not an example of the ``monocultural fallacy'').  

We find that the Sustainability Solution, and the Light Cage Limit specifically, do not adequately address these three issues.  We certainly agree with the general assertion that exponential growth at even modest rates will quickly reach even the highest physical limits.  But it is unclear why reaching such a limit would {\it inevitably} lead to a catastrophic, permanent cultural collapse, rather than slowing of growth.  Further, it is unclear why even such a collapse would preclude the launching of additional colony ships for periods long on a cosmic timescale (as opposed to merely being long on a human timescale).  Finally, and perhaps because the mechanism is unclear, it is unclear why this mechanism would be universal to all colonies of all species, for all time.

The Light Cage Limit hypothesis expressly assumes that a young, rapidly expanding civilization colonizing nearby stars will experience a civilization-wide collapse, via a shared economy.  The Sustainability Solution makes this assumption implicitly by modeling the entire supercivilization as a single unit.   But as we have argued, civilizations spanning multiple stars should be expected to behave more or less independently.

Examples from our own history can provide counterexamples to these sustainability arguments.  Consider, for instance, the first hominid migrations out of Africa, which has many similarities in many ways to stellar colonization.  The independent tribes could interact, but any communication was limited to word-of-mouth (or hand), so there could be no unifying culture and they together formed a loose ``super-tribe'', analogous to our ``supercivilization''.  Hominid tribes at the time had only pedestrian locomotion, and so the super-tribe was effectively more than one ``foot-lifetime'' across.  

Because land was an effectively unlimited resource, the hominid population could grow through the expansion of the number of tribes, regardless of the degree of sustainability of the individual tribes.  Only once all of the most habitable areas accessible by foot had been colonized could resource depletion check global population, even if locally tribes were forced to compete for scarce resources before then.  Analogously, in galactic colonization the relevant resource is {\it other stars}.  So until a galaxy-wide supercivilization is established, there is no resource saturation that would check the growth of the number of civilizations, even if many individual civilizations are unsustainable.  

Put another way, sustainability solutions conflate the issue of the speed of population growth within single civilization, with the speed of the spread of a supercivilization across a galaxy.  Such solutions only solve the Fermi Paradox if all civilizations are so universally unsustainable that none ever manages even a single colony ship; or if long-lived, sustainable alien civilizations for some reason {\it inevitably} and {\it eternally} decline to launch colony ships (which seems to us a hypothesis disconnected from that of ``sustainability'').

\subsubsection{Colonization Need Not Be Driven by Overcrowding or Resource Depletion}

A possible objection to this observation is that without resource depletion or population pressure, there is no motivation to launch colony ships.  In this view, civilizations that are locally slow- or non-growing are also inevitably slow- or non-colonizing.

Indeed, models such as those of \citet{haqq09} and \citet{newman81} often implicitly and sometimes explicitly assume that interstellar colonization is driven by population growth or resource depletion.  This seems to us an unmotivated and dubious assumption.  It seems unlikely that colony ships could transport a sufficient number of souls that they could relieve population pressure, and so colonizing for this purpose makes little sense \citep{vonHoerner75}.  While it is true that such pressures might provide special inducement for small sections of a population to leave for the purposes of avoiding crowding, the very long cruise times to other stars mean that they would typically be leaving to finish their lives on the colony ships.   Given a large population, advanced technology, nearby unpopulated planets, and enough time, colony ships might be launched for any number of other reasons.

As another earth-based counterexample, consider Europeans' spread across the Earth: the initial discovery and colonizations of the Americas or Australia were not undertaken primarily because European population or food scarcity had reached some threshold that inspired mass exploration and emigration.  Rather, they were undertaken by small numbers of people (small relative to their native lands' population) for a variety of social and political reasons.  Further, the primary trigger of the establishment of European colonies was not overcrowding or resource depletion in Europe, but the widespread availability of the technology to cross the oceans and develop self-sustaining colonies on distant shores.  

Analogously, the short maximum galactic colonization time we described in Section~\ref{fastcolony} applies even in cases where resource use is proceeding sustainably and population (and energy use) growth on individual, mature colonies is zero, since it only requires a single colony ship to be launched at least every 10,000 years from every colony, for any reason.   Indeed, {\it sustainability dramatically decreases the colonization time of the galaxy} because it allows for the existence of long-lived civilizations on many stars, each of which can continue colonizing the galaxy for their duration.  

Thus, neither catastrophic social collapse (which should be impossible across a supercivilization anyway) nor slow or zero population growth of mature colonies is inconsistent with Galactic colonization, and so even if the sustainability hypothesis is correct, it should not prevent the colonization of the Galaxy.

Finally, we reiterate that faster-than-light communication and travel provides another way for the sustainability hypothesis to solve the Fermi Paradox, because it allows evasion of the monocultural fallacy.  Such capability allows there to truly be one single galaxy-spanning civilization, capable of coordinating its own sustainability, or its own demise through universal, unsustainable practices.



\subsubsection{Implications for Waste-Heat SETI}

All that said, if the sustainability hypothesis is valid and inevitable, then the average ETI energy supply across a galaxy may be quite low since long-lived civilizations will all have low resource consumption (i.e.\ energy use).  A waste heat search may then need to be extremely sensitive to successfully detect alien waste heat.  In some sense then, even a null waste-heat search will put lower limits on the sustainability of the (undetected) civilizations around the stars and in the galaxy it searches.  The more sensitive the waste heat search is, the lower these limits will be, and the more valid the sustainability hypothesis will appear.

Indeed, the sustainability hypothesis has fascinating implications for this {\it average} energy supply.  Consider one extreme of the sustainability hypothesis: the short-lived, fast growing civilizations.  Since the technology required to send ships to another star (a technology threshold we have nearly reached) is significantly simpler than that of capturing all of a star's energy (we currently short by a factor of $\sim 10^{14}$) one can imagine a scenario where a galactic supercivilization is born, and then some of its component colonies become voracious, developing unsustainable technologies.  They might commonly seek energy supplies larger than their stars can naturally provide, and rather than sustaining themselves for billions of years on starlight, ``burn'' their stars at a higher luminosity (for instance by feeding them to a black hole).  Such civilizations would have very high power supplies (rivaling giant stars, or higher), but be necessarily short-lived (either for ecological reasons or simply because they run out of energy).  Indeed, taken to the extreme, this ``live fast, die young'' scenario results in a galactic supercivilization of short-lived ``energy hogs,'' \citet[the ``graveyard civilization'' of][]{haqq09}.  

These civilizations with short lifetimes, while rare, would also have approximately correspondingly higher waste-heat output, and so still be significant contributors to the galactic heat budget.  A waste-heat search in the integrated starlight of such a galaxy might still succeed: it would be sensitive to a phase average of the civilizations orbiting some significant fraction of the $\sim 10^{11}$ stars in the galaxy, and the ``energy hogs'' might make the average significantly higher than it would otherwise be. 

\subsection{They Exist and We Have Not Noticed Them}

If there is a large hole in Hart's argument it is in his assumption that at least one visitor to the Solar System would have stayed.  Arguing that the Galaxy is filled with ETIs and that none of them colonize nearby stars is an example of the monocultural fallacy; it is much less fallacious to argue that none of the ETIs {\em that has happened to visit the Solar System} has decided to colonize it.   Alternatively, an anthropic argument can be made that our very existence here, pondering this question, is predicated on the Solar System having remained unmolested up to the present day, despite being in a galaxy teeming with ETIs.  In either case the other stars in the Milky Way would host inhabited worlds, and we have simply not yet noticed.  To use an earthly analogy, most individual animals on Earth have probably never seen a human, despite our ability (in principle) to colonize every corner of it.
 
So would we necessarily have noticed that the Galaxy is filled with ETIs?  Given that the age of the Galaxy is consistent with ETIs being billions of years more advanced that we are, we must anticipate that we might not recognize many ETIs if we were to encounter them, any more than an ant colony might recognize humans as a fellow organized species \citep{vonHoerner75}.  This concern must also be applied to any attempts they might make to engineer deliberate signals for us to detect.  \citet{sagan73b} considered the probable timescales involved for the development (and destruction) of ETIs and argued that, even if our Galaxy is saturated with intelligent life, much of it has most likely survived catastrophe, evolved beyond this communication horizon, and is thus invisible to communication SETI.  

One way in which alien civilizations would be very recognizable at any level of technological development would be in their use of energy.  By searching for evidence of energy use, we would free ourselves of the guesswork of communication with unknown intelligence and the assumptions of what form alien life or advanced technology would take, and root ourselves in fundamental science.  

\subsection{The Optimism of Hart's Conclusions}

Hart’s thesis combines the short colonization timescales with the other aspects of his argument to claim that the Milky Way should contain either zero space-faring civilizations, or be filled with them.  Since the latter option appears not to be the case, he concluded that we are alone in the Milky Way.

However, Hart’s arguments apply equally well (or better) to {\it other} galaxies:  every other galaxy represents an essentially independent realization of Hart’s thought experiment. We would therefore expect that all galaxies are either uninhabited by spacefaring civilizations or host (at least one) galaxy-spanning supercivilization.  Indeed, if such galaxy-spanning civilizations could be detected, the relative numbers of inhabited and “sterile” galaxies would then provide a rough guide to the frequency of advanced ETI on a cosmic scale.

Thus, Hart’s pessimism towards the utility of a search for ETI’s in the Milky Way actually translates into an optimism for the existence of galaxy-spanning civilizations in other galaxies.  But if a thorough search for such civilizations in galaxies across the local universe reveals nothing, then one of the following is true:
\begin{itemize} 
\item Galaxy-spanning ETI’s are universally undetectable (perhaps simply because their energy supplies are universally below our detection threshold)
\item Hart’s thesis is fundamentally flawed, or
\item Spacefaring life in the universe is so vanishingly rare that it is effectively unique to Earth.
\end{itemize}

If the first of these is the case, then SETI efforts can proceed by improving our detection thresholds.  If the second of these is the case (perhaps because faster-than-light communication is possible, or interstellar colonization is fundamentally difficult in some way not considered here), a search for within the Milky Way is much more likely to succeed than Hart asserted, and so should be pursued more vigorously.

\section{Past Efforts and the Promise of WISE}

\subsection{The Plausibility and Detectability of Dyson Spheres}

\label{Dyson}

We will argue in Paper II that starlight contains most of the useful free energy in a planetary system, so we expect that any long-lived ETIs with large energy supplies will primarily be driven by stars' energy.  There are two obvious ways to convert a star's mass to useful energy:  by collecting the starlight naturally generated by stellar fusion, and by attempting to extract mass-energy at a faster rate (and perhaps higher efficiency) through something like the Penrose mechanism using a black hole \citep{Penrose}.   Both methods require that the extracted or collected energy be ultimately reradiated as waste heat.

At first glance, it may seem plausible that the complete collection of stellar photons could pose insurmountable engineering problems, but in fact the minimum physical requirements of such an endeavor are not so severe as to be physically impossible.  A Dyson sphere need not be considered as a solid object, and the collection, operating, and radiation surfaces need not be coincident.  Perhaps the simplest model would consist of a swarm of collectors (a ``Dyson swarm'') orbiting very close to a star, converting stellar photons into useful energy (and thus reducing the amount of starlight that escapes).  This energy could then be transmitted (for instance with lasers) to the locations where the alien civilization does its work.  Without specifying their exact nature, we might call these locations ``factories'', or imagine that they are parts of a ``stellar engine'' \citep{badescu2000}.  These factories would radiate the waste heat from the work they do.  

Basic energy balance reveals that these factories would require an outward surface area equal to a sphere 1 AU in radius to radiate a solar luminosity of 255 K waste heat.  Such a structure or series of structures would certainly require mega-engineering of almost inconceivable complexity, but the amount of available mass in a stellar system does not necessarily prohibit such a structure:  even beyond the mass available in the star itself, a single giant planet contains enough material for such a radiating surface (the volume of Jupiter is equivalent to a solid shell 1 AU in radius and over 5 meters in thickness).  The construction of such surfaces, then, requires ``just'' engineering, the desire for more energy use, time, and the application of sufficient energy.  Indeed, a Dyson swarm may not even require central planning, but might arise naturally from many independent subcultures' ever-increasing energy demands.  

We can only imagine (probably unsuccessfully) what kind of projects an ETI could complete with such an energy resource. But regardless of their intentions, thermodynamics dictates that a significant fraction of that energy must be expelled as waste heat corresponding to the operating temperature of their technological efforts. If an ETI prefers terrestrial conditions like ours, this operating temperature would likely then be measured in hundreds of Kelvins, which implies strong radiation in the MIR. Therefore, one promising SETI approach would be to look for the spectral signatures of this process. If the star were perfectly encased by such a shell of structures orbiting at about 1 AU, the resulting spectrum would then appear to be a good, few-hundred-Kelvin blackbody, with a temperature depending on the thermodynamic efficiency \citep{badescu95,slysh85}.

\subsection{Past Searches for Dyson Spheres}

\label{past}

Very wide-field infrared surveys are clearly needed to systematically search for these structures. The Infrared All-Sky Survey ({\it IRAS}; \citealp{IRAS}) provided a hallmark survey for homogeneously investigating infrared sources and identifying Dyson sphere candidates; \citet{slysh85} reasoned that the survey could detect (mostly complete) Dyson spheres out to $\sim$ 1 kpc. The survey immediately showed several potential Dyson sphere candidates, but the infrared colors and low resolution spectra of candidate objects can be confused with stars of many ages obscured by reddened and dusty objects such as stars behind large amounts of extinction, protostars, Mira variables, stars in the AGB phase of evolution and beyond, and planetary nebul\ae.  

Nonetheless, Jugaku et al. performed a series of follow-up searches on catalogs of {\it IRAS} sources, concentrating on those with excess flux around 12 $\mu$ \citep{jugaku91,jugaku95,jugaku97,jugaku00}.  These studies hunted for objects with anomalously red $(K-12\mu)$ colors, using both archival and new $K$-band photometry.  These studies combined determined that there are unlikely any partial to full Dyson spheres around the 365 solar-type stars analyzed within 25 pc. Further results using 2MASS photometry indicated the same with an additional sample of 180 stars within the same distance \citep{jugaku04}.

Later work (\citealp{carrigan09a}; \citealp{carrigan09b}) examined sources observed with the {\it IRAS} low resolution spectrometer (LRS) to determine whether their SEDs could be fit by the blackbody spectra of full or partial Dyson spheres (T = 100--600K). Very few of the 11,000 sources investigated with the LRS were weak Dyson sphere candidates; the spectra of these typically had characteristics of carbon stars with 11.3 mm SiC emission features. The most promising candidate, IRAS 20369+5131, had the most accurate blackbody fit to its spectrum (T = 376K), but is more likely a distant red giant star with no visible 11.3 mm emission peak. \citet{carrigan12} noted that no obvious \HI\ emission of an artificial and intentional origin was detected from this or other Dyson sphere candidates, again underlining the importance of this archeological SETI.  

These source lists do not overlap with another study of Planck spectrum fits to {\it IRAS} data, specifically broadband measurements \citep{timofeev00}. \citet{carrigan09b} argued that fitting LRS data is fundamentally superior to fitting the sparser broadband data, discriminating against filter measurements that only approximately follow a Planck distribution, but have distinct discrepancies in their SEDs. But in any case, an observed deficit of OH and SiO emission for the more promising (however unlikely) sources from any of these programs would be needed to more concretely identify proper Dyson sphere candidates \citep{slysh85} and rule out more customary astrophysical phenomena.

\subsection{Past searches for Galaxy-Spanning Supercivilizations}

Unlike the equivalent searches for these effects on individual stars, few SETI programs to hunt for their accumulated effects of collections of stars have been completed. \citet{annis99b} looked for outliers of the Tully-Fischer relation for a sample of 57 spiral galaxies, and found that the scatter intrinsic to the plots can be attributed to natural causes (e.g., dust extinction) and would only correspond to brightness discrepancies of $\approx$ 50\% or less. He considered magnitude differences of 1.5 mag or more --- stellar optical obscuration by 75\% or more --- as possible candidates for homes of expansive supercivilizations. The same was true for his sample of 106 elliptical galaxies on the fundamental plane. He found no inexplicable outliers on that diagram, either.

Assuming that the emergence of galaxy-wide supercivilizations can be modeled with Poisson statistics, \citet{annis99b} inferred that, without evidence of these supercivilizations in the Milky Way, M31, or M33, the timescale of this evolution could be about 7 billion years, and could rise to hundreds of billions of years if the other galaxies of his sample are considered. This is an important first attempt to evaluate our isolation in the universe based on observable characteristics of nearby galaxies, constraining the number of supercivilizations there may well be in the low-redshift Universe. Of course the assumption that their development is a random process and independent of time is highly uncertain, and seems to contradict, e.g., the time-dependent phase transition arguments of \citet{cirkovic08}.

This type of statistical analysis of galaxy properties, while elegant, has its disadvantages. It is limited to galaxies already discovered and classified in optical catalogs, and so would necessarily miss galaxies with little to no optical emission (i.e., completely obscured by Dyson spheres). \HI\ surveys are starting to uncover \HI\ consolidations like Virgo\HI\ 21\citep{minchin05} that would lie well off either the Tully--Fischer relation or the Fundamental Plane and almost certainly have astrophysical explanations.  Virgo\HI\ 21 in particular has $\approx$ 10$^{8}$ \msun\ of \HI\ spread across 16 kpc, and given its velocity dispersion, has a dynamical mass-to-light ratio of $\approx$ 500 \msun/L$_{\odot}$. There is no presently observed optical component, placing it 12 magnitudes off the Tully-Fischer relation and a clear initial candidate for that methodology. However, Virgo\HI\ 21 is likely a dark halo devoid of a central bright galaxy \citep{minchin05} and not a fully inhabited galaxy. Many unusual galaxies off the traditional relations may well be surrounded by different masses of halos or simply be very dusty. To help eliminate these cases, additional observations and a larger sample size of homogeneously surveyed galaxies (preferably in the infrared) are needed.  

\subsection{The promise of {\it WISE}}

An effective search for ETI's with large energy supplies would image a large sample of stars and galaxies in mid- and far-infrared wavelengths, and look for anomalous colors or SEDs that would indicate the presence of an advanced civilization.   Such a search for ETI's has been performed by \citet{timofeev00}, \citet{jugaku91}, \citet{jugaku95}, \citet{jugaku97}, and \citet{carrigan09b}.  A search for galaxy-spanning civilizations would operate on similar principles, but on the integrated starlight of entire galaxes.

The Infrared Astronomical Satellite \citep[{\it IRAS},][]{IRAS} performed photometry at 12, 25, 60, and 100 $\mu$ and low resolution spectroscopy from 7.5 to 23 $\mu$. IRAS offered sensitivity of around 0.2 Jy (1 Jy at 100$\mu$) \citep{IRASFSC}, and angular resolution of 30$^{\prime\prime}$ at 12$\mu$, and 2$^\prime$ at 100 $\mu$.  This has been the primary dataset used in previous searches for alien waste heat.

High hopes for the sensitivity of {\it IRAS} to Dyson spheres were dashed by the unexpected discovery of the infrared cirrus \citep{IRcirrus}.  The high backgrounds made the point source sensitivity of the relatively large {\it IRAS} beam significantly lower than anticipated at 60 and 100 $\mu$, as did the corresponding PAH emission in the shorter-wavelength bands.

The High Frequency Instrument of the {\it Planck} mission \citep{Planck} detected galaxies at 350 $\mu$  as compact sources with 4\farcmin5 resolution and a sensitivity of 3 Jy, and might be used for the (non-)detection of far-infrared (FIR) thermal emission from dust from bright sources (to rule out extraterrestrial origin of the MIR emission).

In 2009, the Wide-field Infrared Survey Explorer \citep[{\it WISE},][]{WISE} launched and began its all-sky survey.  \WISE\ provides photometry at 3.4, 4.62 12 and 22 $\mu$ (the $W1$,$W2$,$W3$, and $W4$ bands), with angular resolution of 6\farcs1, 6\farcs4, 6\farcs5, and 12\arcs, and sensitivities of 0.07, 0.1, 0.9, and 5.4 mJy, respectively  (with $\sim 3\%$ photometric accuracy) \citep{WISEexp,WISEphot}  \WISE\ thus provides 5 times better angular resolution and  1000 times better sensitivity than {\it IRAS}.  {\it WISE} thus affords us a deeper and more accurate glimpse into our own Galaxy for finding evidence of Dyson spheres, and a significantly improved ability to accurately detect, resolve, and image approximately $10^5$ galaxies across the sky in four infrared photometric bands and hunt for galaxy-spanning civilizations among them.

Additional surveys, such as the SWIRE stare, GLIMPSE, and MIPSGAL of the {\it Spitzer} Space Telescope, provide superior sensitivity and resolution to \WISE, albeit with more limited spatial coverage.  Together, these surveys provide a new and unprecedented glimpse into the universe of reprocessed starlight, including, potentially, the starlight that has been repurposed by ETI's.

\section{Results of a Thorough Waste-Heat Search}

\label{results}

\subsection{The possibility and consequence of a dispositive null detection}

The primary advantage to a waste-heat search for alien civilizations is that it makes no assumptions about how or whether an alien civilization would attempt communication.  Further, since it can be extended to searches for supercivilizations in other galaxies, it can address resolutions to the Fermi-Hart paradox that allow for the ubiquity of spacefaring ETI's in the universe but excuse our lack of detection (for instance, many ``sociological'' explanations for their unwillingness to contact us).  

This is in contrast to contact SETI, where the number of frequencies, duty cycles, and transmission methods of ETIs is large, and potentially inexhaustible.  A null result in contact SETI is thus a failure to detect a certain class of ETI communication, while a thorough search for waste heat that comes up empty has the potential to be a successful demonstration of the absence of large ETI energy supplies with few hundred Kelvin waste heat temperatures. 

If the number of ETIs per galaxy is as high as optimistic solutions to the Drake equation suggest \citep{DrakeEq,Papagiannis80}, then it is plausible that some fraction of galaxies will show evidence of large scale civilizations.  

A null detection of galaxy-spanning ETI's with waste heat luminosities above some upper limit in a large sample of galaxies would thus correspond to an upper limit on alien energy supplies in general, and individual upper limits on every galaxy in the sample.  For elliptical galaxies, this upper limit might be similar to current limits for nearby stars (of order a few percent).  

This would itself be an interesting result; to wit, at least one of the following must be true about galaxy-spanning alien civilizations in the search volume:

\begin{itemize}
\item They do not exist, or are sufficiently rare that they are not in our search volume
\item They exist, but their total energy supplies are {\it universally} below the search's detection threshold
\item ETI's with large energy supplies {\it universally} expel waste heat at low temperatures (i.e.\ wavelengths longer than the capabilities of the search), perhaps because their energy supply is not starlight (see Paper II)
\item Spacefaring ETI's {\it inevitably} discover and universally employ physics that makes their civilizations effectively invisible in the MIR despite having large energy supplies (for example, expelling their waste heat as neutrinos, efficiently using their energy supply to emit low-entropy radiation, employing energy-to-mass conversion on a massive scale, or by violating conservation of energy).
\end{itemize}
 
Indeed, since MIR-bright ETI's with energy supplies in excess of their host galaxy's luminosity would be all but trivial for a survey such as \WISE\ to detect and distinguish out to great distances (it would resolve them out to $z\sim0.1$), the absence of any such galaxies would effectively rule out the possibility that physics allows for an easily tapped source of ``free'' energy (e.g.\ zero-point energy) at high effective temperatures, if galaxy-spanning ETI's exist (and such a possibility would make it significantly easier for them to exist in the first place).  The absence of such civilizations would thus rule out an entire class of either exotic physics or ETI's.  We quantify these statements in Paper II.

The last two or three options on our list have little overlap with the resolutions to the ``contact'' version of the Fermi paradox, demonstrating the complementary nature of waste-heat SETI with communication SETI.  To wit, our uniqueness as an intelligent, spacefaring civilization would explain the null detections in both contact SETI and artifact SETI, but most of the other options on this list would not.  The most parsimonious explanation for a null result from both methods, then, would simply be that we are alone in the Universe.  Other explanations would need to invoke multiple reasons for our failure to detect ETI's.

Another perspective is that a null detection of galaxy-spanning ETI's in the Local Universe is in conflict with Hart's thesis that spacefaring civilizations will quickly colonize their galaxy.  Unless humanity is alone in the {\it Universe}, such a null detection would lend evidence to the hypothesis that spacefaring life rarely spreads far beyond its home star system, in which case searches within the Milky Way are much more likely to succeed than Hart argued.

\pagebreak
\subsection{Relation to the Rare Earth hypothesis}

The Rare Earth hypothesis of \citet{Ward00} states that ``primitive'' life is common in the universe but that {\it animal} life is unique to Earth.  While null detections of waste-heat from ETIs would be consistent with this hypothesis, we stress that, unlike \citeauthor{Ward00}, we have not assumed that ETIs need be animal.  Any spacefaring agent that satisfies our ``physicist's'' definition of intelligent life should generate detectable waste heat.

So, while a null detection of artifact SETI supports the conclusion of the Rare Earth hypothesis, it does not necessarily support the reasoning and evidence presented by \citeauthor{Ward00} in support of it.

\subsection{The possibility and consequence of a positive detection}

A positive detection of waste heat from an alien civilization would have several consequences.  Beyond demonstrating that alien life exists, it would also allow us to essentially ``peek ahead'' at the nature of engineering of a vastly more advanced species, and perhaps allow us to test our current theories of fundamental physics.  Such a detection would:

\begin{itemize}
\item Demonstrate that there are no insurmountable obstacles to energy supplies comparable to the luminosities of stars or galaxies
\item Demonstrate that conservation of energy and the laws of thermodynamics, as we understand them, are not circumvented on a large scale by at least some very advanced civilizations, and so may be absolute.  This might imply that attempting to overcome them at our level of technology would be fruitless.
\item Allow analysis of the extent of the alien supercivilization, and so  a measure of characteristic travel speed, potentially revealing the practicalities of travel near the speed of light.
\end{itemize}

In short, it would validate the assumptions that underlie the search for alien waste heat.

The fraction and extent of obscuration and reradiation by artificial constructions would certainly highlight important constraints on the unknown engineering of individual Dyson spheres, e.g., the fraction of stellar coverage and operating temperatures. They would also point to the colonization habits of ETI that have been thus far so difficult to anticipate, e.g., the extent, patterns, and timescales of their conquests.

\section{Conclusions}

\label{conclusions}

Artifact SETI complements communication SETI because it can help narrow and focus searches for communication on targets with unusually high MIR emission, and because artifact SETI alone will be hard pressed to prove that an unusual source is artificial.  Together, the two methods may achieve what either one alone has not.
  
Hart's conclusion was that if his thesis was correct, ``({\sc i}) an extensive search for radio messages from other civilizations is probably a waste of time and money; and ({\sc ii}) in the long run, cultures descended from ours will probably occupy most of the habitable planets in our Galaxy.''  We do not entirely agree.

Regarding our long-run future if the Milk Way is ``empty'', Hart's argument that destiny of humanity is that of a ubiquitous Galactic diaspora appears to be robust, {\it provided} we can establish self-sufficient colonies over a sufficiently large volume that our species' extinction becomes effectively impossible.  Until then {\it homo sapiens} is vulnerable to global catastrophe (including self-induced annihilation) and this destiny is uncertain.  It has been noted many times before that humankind is the first species capable of deliberately averting its own destruction (and, too, causing it); this underscores Sagan's assertion that Earth,  our ``Pale Blue Dot'', ``is where we make our stand'' \citep{PaleBlueDot}.  Our species thus appears to be at the tipping point between eventual extinction and immortality, which only heightens the stakes of stewardship of our species and our biosphere.

That said, we find that this observation does little to help ``solve'' the Fermi Paradox.  We find that solutions that invoke sustainability suffer from variations on what we dub the ``monocultural fallacy'' (i.e., the attribution of certain traits or tendencies across all species, across all civilizations within a species, and across the entire duration of a civilization).  We also find that some such solutions conflate the growth of populations {\it within} a civilization with the growth in the {\it number} of civilizations in a widely separated ``supercivilization'' (by which we mean civilizations of common origin but separated by light travel times long compared to the timescales of cultural evolution).  We also find that these solutions are weakened by our observation that the {\it maximum} time to colonize the Galaxy for a spacefaring civilization is likely to be quite short --- on the order of a few galactic rotations.

Hart's conclusion that we are alone in the Milky Way is certainly the simplest and perhaps most obvious explanation for the ``Great Silence,''  but speculation about the nature of extraterrestrial life, especially life-as-we-do-not-know-it, is inherently so fraught with ``unknown unknowns'' that  declaring any well-considered SETI effort ``a waste'' is to commit a failure of imagination.\footnote{To borrow a phrase from \citet{Clarke3}.}   To justify SETI it is sufficient to argue that the odds are uncertain, the payoff for success high, and the cost of searching low.  

But Hart's apparently pessimistic conclusion actually quite optimistic and allows for a superior justification: it implies that unless humanity is alone in the {\it local universe}, there should exist civilizations spanning many of the galaxies we see.  If Hart's conclusions are correct, then even though any search for ETI's within the Milky Way should fail, {\it unless spacefaring life is unique to Earth, an extragalactic search for has a sufficiently good chance of ultimate success that searches are worthwhile and should be developed}.  Contrariwise, if Hart's thesis is incorrect, then a search for galaxy-spanning civilizations in {\it other} galaxies might be futile (because they may not exist), but a search for circumstellar civilizations within the Milky Way is more likely to bear fruit than Hart asserted.  Searching for both types of civilizations, with both artifact and communication SETI, is thus a compelling path forward for SETI efforts.

We have provided a ``physicist's definition'' of life and intelligence, and combined this with Hart's argument to argue that energy supply and waste heat are useful diagnostic of large alien energy supplies.  Past efforts to search for waste heat have been hampered by the lack of a sensitive, wide-field mid-infrared survey.  Today, \WISE\ and {\it Spitzer} allow us to search for Galactic and extragalactic alien civilizations with large energy supplies, and to put some of Hart's conclusions to an empirical test.

A successful detection of a distant alien civilization will give us a ``peek'' into the limitations and possibilities of advanced ETIs.  A null detection would implicate either the violation by alien technology of physical laws we consider fundamental, or the non-existence of ETI's with large energy supplies and mid-infrared waste heat in our search volume.

\acknowledgements

This research is supported entirely by the John Templeton Foundation through its New Frontiers in Astronomy and Cosmology, administered by Don York of the University of Chicago.  We are grateful for the opportunity provided by this grant to do this research.

The Center for Exoplanets and Habitable Worlds is supported by the Pennsylvania State University, the Eberly College of Science, and the Pennsylvania Space Grant Consortium.

This publication makes use of NASA's Astrophysics Data System.  

The literature on the spread, likelihood, and form of extraterrestrial intelligence in the universe is sprawling.  This paper is not intended to be a comprehensive review of the problem of extraterrestrial civilizations.  While we have tried in this and the later papers of this series to make it clear which contributions to the discussion are purely ours in the text, and while we have strived to provide appropriate citations to unoriginal ideas, it can be difficult to provide, or even recognize, proper citations for many important ideas pertaining to ETIs.   In many cases, we may not even be aware of our influences. 

Many important ideas and seminal contributions appearing outside of peer-reviewed journals, for instance in conferences, reports, white papers, and even science fiction books, teleplays, and film, making them tricky to track down and cite.  To give just three examples:  The analogy of ants as an inferior species incapable of even comprehending humanity for our incomprehension of ETIs could cite ``The Search'' by Randall Munroe\footnote{http://xkcd.com/638/}, but an earlier citation to ``Mind War'' by J.\ Michael Straczynski (1994 {\it Babylon 5}, Season 1, Episode 6, Act 5) is at least as appropriate; just before publication we discovered an early instance in \citet{vonHoerner75}, but even earlier attestations may exist.   The ``grey goo'' scenario apparently originates with Eric Drexler's novel {\it Engines of Creation}, but the full implications were fleshed out later, most notably in an essay by Robert A.\ Freitas Jr.\ for which we provide a URL in footnote \ref{goo}.  We have cited \citet{vonHoerner75} for the SETI implications of the ultimate Malthusian limit of waste heat, but the online discussion surrounding Tom Murphy's blog posts\footnote{http://physics.ucsd.edu/do-the-math/2011/07/galactic-scale-energy/} on ``Galactic-Scale Energy'' deserves citation, as well.  
 
Even restricting oneself to the refereed literature, it can be difficult or impossible to find the origins of ideas that may have crept into popular culture, or to be sure an idea has not been thoroughly discussed somewhere.   For instance, the term ``supercivilization'' appears in \citet{kardashev85} but likely has earlier attestations.  In short, we apologize that the references included in this series of papers are not comprehensive, and we acknowledge that many ideas not original to us likely appear without citation.

We thank Jill Tarter, Freeman Dyson, Franck Marchis, Bill Cochran, Marshall Perrin, and Geoff Marcy for helpful contributions and discussions, and Roger Griffith for his work with us on the \WISE\ data set.  We thank our referees for appropriately challenging our conclusions and sharpening our reasoning.  

S.S.\ thanks David Brin for many interesting discussions over the years.  J.T.W.\ thanks Jill Tarter, Dan Wertheimer, and most especially Geoff Marcy for teaching him, by outstanding example, and that searches for ETI's can and should be a scientifically respectable endeavor.




 \clearpage

\end{document}